\documentclass{emulateapj}

\begin{document}
\slugcomment{accepted \apj}

\title{The Size Scale of Star Clusters}


\author{Juan P. Madrid, Jarrod R. Hurley, Anna C. Sippel}

\affil{Center for Astrophysics and Supercomputing, Swinburne
University of Technology, Hawthorn, VIC 3122, Australia}


\begin{abstract}

Direct $N$-body simulations of star clusters in a realistic Milky Way-like 
potential are carried out using the code \texttt{NBODY6}. Based on these simulations
a new relationship between scale size and galactocentric distance is derived:
the scale size of star clusters is proportional to the hyperbolic tangent of
the galactocentric distance. The half-mass radius of star clusters increases 
systematically with galactocentric distance but levels off when star clusters 
orbit the galaxy beyond $\sim$40 kpc. These simulations show that the half-mass 
radius of individual star clusters varies significantly as they evolve over a 
Hubble time, more so for clusters with shorter relaxation times, and remains 
constant through several relaxation times only in certain situations when 
expansion driven by the internal dynamics of the star cluster and the 
influence of the host galaxy tidal field balance each other.
Indeed, the radius of a star cluster  evolving within the inner 20 kpc of a realistic 
galactic gravitational potential is severely truncated by tidal interactions and does not 
remain constant over a Hubble time. Furthermore, the half-mass radius of star clusters 
measured with present day observations bears no memory of the original cluster size. 
Stellar evolution and tidal stripping are the two competing physical mechanisms that 
determine the present day size of globular clusters. These simulations also show that 
extended star clusters can form at large galactocentric distances while remaining fully 
bound to the host galaxy. There is thus no need to invoke accretion from an 
external galaxy to explain the presence of extended clusters at large 
galactocentric distances in a Milky Way-type galaxy. 

\end{abstract}


\maketitle

\section{Introduction}

Star clusters are an increasingly diverse family. During the last decade 
the discovery of stellar systems brighter, larger, and more massive than the 
``standard" star cluster has blurred the distinction between globular 
clusters and dwarf galaxies. Ultra-compact dwarfs (UCDs) are the prime 
example of stellar systems with physical parameters between those of 
globular clusters and dwarf elliptical galaxies (Hilker et al.\ 1999;
Drinkwater et al.\ 2000). At the other end of the luminosity range, 
faint and extended star clusters have also been discovered, 
notably as satellites of the Andromeda galaxy (Huxor et al.\ 2005; 
Mackey et al.\ 2006) and NGC 1023 (Larsen \& Brodie 2000). 

These newly discovered stellar systems have bridged the gap in physical size
thought to exist between globular clusters and compact elliptical galaxies 
(Gilmore et al.\ 2007). The well defined linear relations between physical size 
and total magnitude for galaxies with masses greater than $10^8 M_{\odot}$
can be extrapolated to UCDs. However, dwarf galaxies and star clusters form two 
branches in the size-magnitude plane where these two physical parameters 
are uncorrelated (Misgeld \& Hilker 2011 their Figure 1; see also McLaughlin 2000). 

In recent years, studies based on the relatively wide field of view and 
superb resolution of the Advanced Camera for Surveys onboard the {\it Hubble 
Space Telescope (HST)}, have found a strikingly constant median effective radius, 
or equivalently half-light radius, for extragalactic globular clusters of $<r_h>\sim 3$ pc. 
Jord\'an et al.\ (2005) provide a prime example of such work as they accurately 
determined the structural parameters of thousands of globular clusters 
associated with 100 early type galaxies of the Virgo cluster (ACS Virgo Cluster Survey). 
Masters et al.\ (2010) replicated this work for 43 galaxies in the Fornax cluster 
(ACS Fornax Cluster Survey). One of the main findings of these papers is that 
thousands of star clusters spanning more than four magnitudes in luminosity have 
the same median value of $<r_h> \sim 3$ pc.

Why is the size distribution of star clusters narrowly centered around three 
parsecs? Why do only a few clusters become extended with effective radii of 
ten parsecs or more? In this work, advanced $N$-body models are carried out with the aim 
of determining the most important physical mechanisms that mold the 
characteristic radii of star clusters. The impact of the host galaxy tidal 
field on the size of orbiting star clusters is probed in detail 
by evolving several models at different galactocentric distances ($R_{GC}$).

The empirical qualitative dependence between size and galactocentric distance of 
star clusters has been clearly established in several observational studies 
beginning with the work of Hodge (1960, 1962). The $N$-body models that have 
been performed allow us to quantify the influence of the tidal field, generated
by a Milky-Way or M31 type galaxy, on satellite star clusters. We thus determine a new
relation between the scale sizes of star clusters and galactocentric distance.
This relation is a proxy for the host galaxy gravitational potential.


\section{The models}

All $N$-body simulations are performed using the \texttt{NBODY6} code (Aarseth 1999; 
Aarseth 2003). This code performs a direct integration of the equations of motion for all 
{\it N} stars and binaries in a star cluster and includes a comprehensive 
treatment of stellar evolution (Hurley et al. 2000, 2005).  This code also includes
a detailed handling of binaries accounting for close encounters, mergers, and 
the formation of three and four body systems (Tout 2008; Hurley 2008; Mardling 2008;
Mikkola 2008).

The version of \texttt{NBODY6} that is used in this work was specially modified 
to run on a Graphic Processing Unit (GPU: Nitadori \& Aarseth 2012). 
During the last decade the clock rates of Central Processing Units (CPUs) 
have been practically stagnant while GPUs provide a proven alternative 
for high-performance computing (Barsdell et al.\ 2010), and 
particularly for $N$-body codes (Hamada et al.\ 2009). 
The models are run at the Center for Astrophysics and Supercomputing of 
Swinburne University. The GPUs in use are NVIDIA Tesla S1070 cards.
Earlier versions of this code (i.e.\ \texttt{NBODY4}) were run on special-purpose
GRAPE hardware (Makino et al.\ 2003) but the performance of the GPU 
version of this code is comparable or superior to previous efforts to 
improve computing time. The calculations are carried out in $N$-body units, 
i.e.\ $G=1$ and $-4E_0=1$, where $E_0$ is the initial energy (Heggie \& Mathieu 1986). 
The results are scaled back to physical units once the computation of the models is complete.


\subsection{Numerical Simulations Set Up}

The initial set up for the simulations carried out here is similar to the 
work of Hurley \& Mackey (2010). All simulations have an initial number of 
particles of $N=10^5$. Of these, 5\% are primordial binary systems, that is,
95 000 single stars and 5000 binary systems. The most massive star has a mass 
of $M_{max}=50 M_{\odot}$ while the minimum mass for a star is $M_{min}=0.1 M_{\odot}$. 
The initial mass distribution for all stars follows the stellar Initial  
Mass Function (IMF) of Kroupa et al.\ (1993). The models start off with a total initial 
mass of $M_{tot}\approx6.3 \times 10^4 M_{\odot}$. The $N=10^5$ stellar systems have 
the initial spatial distribution of a Plummer sphere (Plummer 1911) and 
an initial velocity distribution that assumes virial equilibrium. The modeled 
clusters stars have a metallicity of $Z=0.001$ or $[Fe/H] \approx -1.3$. 

These simulations start after the clusters will have undergone 
expansion as the result of the removal of residual gas left over from
star formation. Thus, at $t=0$ all stars are assumed to have formed and 
be on the zero age main sequence, with no gas present. The initial
three-dimensional half-mass radius is 6.2 pc. 

The initial set up for each simulation is exactly the same with the exception
of the initial galactocentric distance. Individual simulations of star 
clusters evolving on circular orbits at different galactocentric distances
were obtained, i.e. $R_{GC}=$ 4, 6, 8, 8.5, 10, 20, 50 and 100 kpc.
The initial plane of motion of the star clusters is 22.5 degrees 
from the plane defined by the disk.


\subsection{Galactic Tidal Field Model}

Earlier models from $N$-body simulations have made the simplifying 
assumption that the gravitational potential of the host galaxy can 
be represented by a central point mass only (e.g.\ Vesperini \& Heggie 
1997; Baumgardt 2001; Hurley \& Bekki 2008). The version of \texttt{NBODY6} 
used here models the bulge as a point source but also includes a halo 
and a disk as components of the Milky-Way galaxy and its gravitational potential. 
This more realistic implementation of the host galaxy potential has been
incorporated in recent work (e.g.\ K\"upper et al.\ 2011).

To model the disk \texttt{NBODY6} follows Miyamoto \& Nagai (1975) whom 
combined the potential of a spherical system (Plummer 1911) and the potential 
of a disk-like mass distribution (Toomre 1963) into a generalized analytical
function that elegantly describes the disk of spiral galaxies:

\begin{equation}
\Phi(r,z)=\frac{GM}{\sqrt{r^2 +[a+\sqrt{(z^2+b^2)}]^2}}.
\label{eqn:eqmm}
\end{equation}

Here $a$ is the disk scale length, $b$ is the disk scale height, 
$G$ is the gravitational constant, and $M$ is the mass of the disk component. 
The elegance of this formulation resides in the fact that it can model
both an infinitely thin disk or a sphere by varying the two scale
factors $a$ and $b$. The values used here are $a=4$ kpc for the disk scale length, 
and $b=0.5$ kpc for the vertical length (Read et al.\ 2006). Formally, 
the Miyamoto \& Nagai disk extends to infinity. However, the strength 
of the disk potential asymptotes towards zero at large radii: 
with $a=4$ kpc and $b=0.5$ kpc, the density at 40 kpc drops to 0.1\% 
of the central value.

The masses of the bulge and disk are $1.5\times10^{10}M_{\odot}$ and $5\times10^{10}M_{\odot}$,
respectively (Xue et al.\ 2008). The galactic halo is modeled by a logarithmic 
potential. When combined together, the potential of the halo, disk, and bulge are constrained 
to give a rotational velocity of 220 km/s at 8.5 kpc from the galactic center (Aarseth 2003). 
A detailed discussion of the galactic tidal field model is given by Praagman et al.\ (2010). 

A tidal radius for a star cluster in a circular orbit about a point-mass galaxy, 
initially postulated by von Hoerner (1957), can be approximated by the King (1962) 
formulation:

\begin{equation}
r_{\rm t} = \left( \frac{M_{\rm C}}{3M_{\rm G}} \right)^{1/3}R_{\rm GC}
\label{eqn:eqmm}
\end{equation}

where $M_{\rm C}$ is the mass of the cluster, and $M_{\rm G}$ is the mass of the 
galaxy. The equivalent expression for a circular orbit in the Milky-Way potential 
described above is:

\begin{equation}
r_{\rm t} \simeq \left( \frac{GM_{\rm C}}{2\Omega^2} \right)^{1/3}
\label{eqn:eqkk}
\end{equation}

where $\Omega$ is the angular velocity of the cluster (K\"upper et al.\ 2010a).
The tidal radius is where a star will feel an equal gravitational pull 
towards the cluster and towards the galaxy center in the opposite direction. 
A detailed study of the tidal radius of a star cluster for different galaxy 
potentials is given by Renaud et al.\ (2011). 

In the $N$-body code the gravitational forces owing to {\it both} the cluster 
stars and the  galaxy potential are taken into account for all stars in the 
simulation, independently of the definition of the tidal radius. However it is 
common practice to define an escape radius, beyond which 
stars are deemed to be no longer significant in terms of the cluster potential
and thus have no further input on the cluster evolution. Stars are removed from 
the simulation when they fulfill two conditions: (i) their distance from the 
cluster center exceeds the escape radius, and (ii) they have positive energy when 
the external field contribution is taken into account.

In our simulations we have estimated the escape radius as twice the tidal
radius given by Eq.\ (2) with $M_G=10^{11} M_{\odot}$. At all times this
gives a value in excess of the tidal radius given by Eq.\ (3). When presenting
results, we only consider stars that lie within the tidal radius given by
Eq.\ (3).

\smallskip
\section{Previous work}

A pioneering study of this subject was carried out by Vesperini \& Heggie (1997) 
who investigated the effects of dynamical evolution on the 
mass function of globular clusters through simulations based on \texttt{NBODY4}. 
This was the most up-to-date version of the code at the time of their work and 
only included a basic treatment of stellar evolution. Also, 
the total number of particles that Vesperini \& Heggie (1997) were able 
to simulate was limited to $N = 4096$ due to restrictions imposed by the 
hardware. Under the conditions used here a simulated cluster with only $N=4096$ 
orbiting at $R_{GC}$= 8 kpc is dissolved in 1.1 Gyr.  
Despite the computational limitations to which Vesperini \& Heggie (1997) were 
subject, they set an inspiring precedent for this work. Particularly relevant is the trend 
of increasing mass-loss with decreasing galactocentric distance, albeit with 
the galaxy potential modeled as a point-mass.

Baumgardt \& Makino (2003)  studied the stellar mass function of star clusters using
\texttt{NBODY4} but with more realistic particle numbers, going up to $N=131072$.
They determined that stellar evolution accounts for $\sim 1/3$ of the total 
mass lost by the cluster. The upper mass limit of the stellar initial mass function
used by Baumgardt \& Makino (2003) was $M_{max}= 15 M_{\odot}$, while 
in the simulations presented here $M_{max}= 50 M_{\odot}$. Baumgardt \& Makino (2003) 
found that owing to mass segregation low-mass stars are prone to be depleted from the 
star cluster. This depletion of low mass stars inverts the slope of the IMF. Unlike 
Baumgardt \& Makino (2003) the models presented here include a population of 5000 
primordial binary systems. For the original spatial distribution Baumgardt \& Makino 
(2003) used a King profile while here a Plummer sphere is used. 


\section{Total mass}

\texttt{NBODY6} readily yields fundamental numerical parameters of 
the simulated star clusters such as the number of stars, number of binaries, 
total mass, core mass, half-mass radius, relaxation time, and velocity 
dispersion, to name a few.

Star clusters evolving in a galactic potential lose mass due to stellar 
evolution, two body relaxation, few-body encounters, and tidal interactions 
with the host galaxy. In the following sections the two main phases of 
mass-loss are discussed.


\subsection{Initial Stellar Evolution}

During the first Gyr of evolution star clusters lose large amounts of mass
owing to stellar evolution. Massive stars with M$\sim 10 M_{\odot}$ and above 
such as OB stars, luminous blue variables, and Wolf-Rayet stars have large 
mass-loss rates of the order of $\dot{M}\sim 10^{-5} M_{\odot}$ per year (Vanbeveren et al.\ 1998;
Shara et al.\ 2009). 
The mass loss rates of these stars is the dominant factor of their evolution. 
In fact, mass-loss determines whether a massive star becomes a supernovae or a 
long duration gamma-ray bust (Vink et al.\ 2000, Shara et al.\ 2009). 
Massive stars are short lived, for instance Wolf-Rayet stars with initial 
masses greater than $20M_{\odot}$ are expected to survive only 10
Myr (Hurley et al.\ 2000). 

The $N$-body simulations plotted in Figure 1 show that star clusters, 
independently of galactocentric distance, lose $\sim 1/3$ of their mass during 
the first Gyr of evolution. This is in agreement with the findings of 
Baumgardt \& Makino (2003). The stellar evolution of massive stars 
discussed above is responsible for the mass loss that triggers an expansion 
of the cluster due to a reduced gravitational pull. The impact of mass loss 
on the scale size of a star cluster is discussed in the next section.

\subsection{Two phases of mass-loss}

The two main regimes of mass loss evident in these simulations are mass loss from 
stellar evolution and mass loss due to tidal stripping. The former initially increases
the cluster size, while the latter decreases the cluster size. Whether or not either
of the two mechanisms dominates will determine the size of the star 
cluster (Gieles et al.\ 2011). 

As mentioned above, stellar evolution mass-loss occurs on a 
rapid timescale at early times. It subsequently slows owing to the longer evolution 
timescales of low-mass stars but maintains a steady presence throughout the ongoing life 
of a cluster. Mass-loss resulting from tidal stripping is linked to the two-body relaxation 
timescale of the cluster: two-body interactions gradually increase the velocities of the low-mass 
stars, pushing them to the outer regions where they are lost across the tidal boundary 
in a process that is often called evaporation (McLaughlin \& Fall 2008). 
For star clusters evolving at small galactocentric distances
(e.g. simulations at 4 and 6 kpc) the dominant mechanism of mass loss is tidal 
stripping. These clusters have relatively short two-body relaxation timescales. 
In contrast, for clusters evolving at large galactocentric distances 
(with longer relaxation timescales) the 
dominant mass loss mechanism is stellar evolution. It is clear from Figure 1 that
the closer a star cluster orbits to the center of the galaxy, the more accentuated its
mass loss is. Figure 1 shows how the mass-loss rate evolves towards an asymptotic 
linear behavior after a Hubble time, in agreement with Baumgardt \& Makino (2003).
Few-body interactions can also lead to the loss of stars through ejections. 
This process can cause small-scale fluctuations in cluster size on short timescales.

 \begin{figure} 
 \plotone{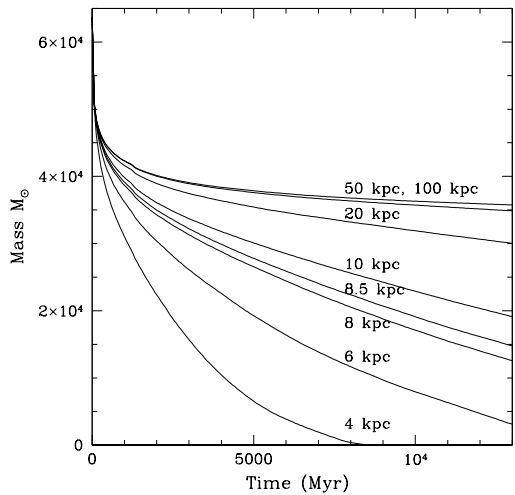} 
 \caption{Mass loss of simulated star clusters at different galactocentric distances.
 Note how the simulated star cluster evolving at 4 kpc has lost all of its mass after 8.4 Gyr.
 \label{fig1}}
 \end{figure}

\section{Characteristic radii}

Different characteristic radii are commonly used for star clusters.
Observers obtain two dimensional half-light radii to high accuracy, 
but are  limited in their choice of characteristic radii. 
For simulated clusters, different scale size values are readily 
available such as the half-mass radius, $r_{hm}$, or core radius. Radii of 
simulated clusters can be expressed in two or three dimensions, 
the latter is adopted in this work. Hurley \& Mackey (2010) show that 
the value of the half-{\it mass} radius derived with \texttt{NBODY6} is on 
average 1.6 times the value of an observed half-{\it light} radius. Moreover,
different characteristic radii describing the core of star clusters
are known to evolve in a self-similar fashion (Baumgardt et al. 2004).

The definition of core radius is different in observational and 
theoretical studies. In observational studies the core radius is 
generally defined as the radius where the surface brightness falls 
to half its central value (King 1962). In our simulations, the core 
radius is a density-weighted average distance of each star to the 
point of highest stellar density within the cluster (Aarseth 2003). 


\subsection{Observed and primordial half-mass radius}

The evolution of the half-mass radius of simulated star clusters with time 
is plotted in Figure 2. Clusters start  with an initial half-mass 
radius of $\sim$~6.2 pc and undergo an expansion 
triggered by stellar evolution within the first Gyr as 
discussed above. An initial half-mass radius of $\sim$~6.2 pc might 
seem larger than average but after a Hubble time of evolution King (1962) models
fitted to the light profiles of our models yield two dimensional half-light radii that are 
consistent with observations, i.e.\ around 3 pc (Sippel et al.\ 2012).


In their original work, Spitzer \& Thuan (1972) found that the effective radius
of an isolated star cluster remains constant through several relaxation times. 
Their important result is confirmed in Figure 2 but only for the model
evolving at 20 kpc from the galactic center. In this particular model
the star cluster experiences only weak tidal interactions with the host 
galaxy. After the initial phase of rapid expansion driven 
by stellar evolution, the longer-term steady expansion driven by the internal dynamics 
is balanced by the presence of the tidal field. Thus,
the half-mass radius of this simulated star cluster 
remains constant over the last eight Gyr of evolution in agreement 
with Spitzer \& Thuan (1972). 



 \begin{figure} 
 \plotone{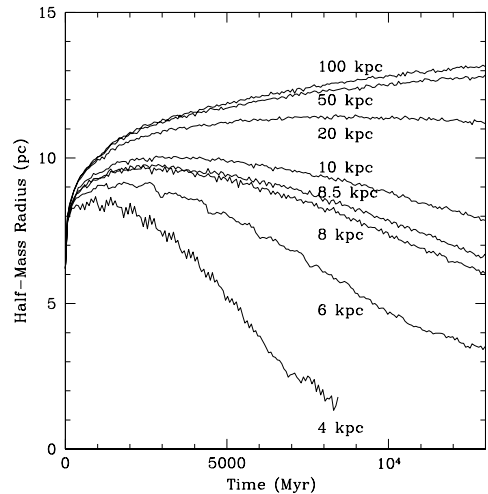} 
 \caption{Evolution of the 3D half-mass radius of simulated star clusters at different 
 galactocentric distances. A simulated star cluster evolving at 4 kpc from the 
 galactic center dissolves before a Hubble time. Simulations of star clusters 
 orbiting the galaxy at less than 20 kpc experience a truncation of their half-mass
 radius due to tidal interactions. 
 \label{fig2}}
 \end{figure}


More importantly, the effective radii of all star clusters evolving at 
$R_{GC} < 20$ kpc are significantly affected by tidal interactions that 
truncate the cluster size during its evolution. 
The results of Spitzer \& Thuan (1972) are often misquoted in observational 
studies as proof that the effective radius remains constant over many relaxation 
times and that the $r_h$ measured today is a faithful tracer of the original
size of the proto-cluster cloud.

As shown in Figure 2, the half-mass radii of simulated clusters 
evolving at 50 and 100 kpc are always increasing. For 
these two simulations, in virtual tidal isolation and with long relaxation times,
their half-mass radii reach about twice the value of the initial half-mass radius,
i.e.\ $r_{hm}/r_{hm0}\approx2$ after a Hubble time (Gieles et al.\ 2010).

Most observational studies of extragalactic star clusters aimed at determining 
their scale sizes are carried out with the {\it Hubble Space Telescope} due to 
its unique resolution (e.g.\ the ACS Virgo Cluster Survey by Jord\'an et al.\ 2005). 
The drawback of space-based detectors is the small field of view that only covers, in general, 
a physical scale of a few kiloparsecs in radius surrounding the core of the host galaxy. 
As shown in Figure 2 the inner ten kiloparsercs of a galaxy is where the tidal field
has the strongest impact on the size of the satellite star cluster, with a dependence 
on galaxy mass. The effective radius of a star cluster evolving in a realistic galaxy 
potential loses any memory of its earlier values. As shown in Figure 2, only the 
simulated star cluster evolving at 20 kpc has an effective radius that remains constant
through several relaxation times which is a result of its particular circumstances, being in 
a position within the galaxy where internally driven expansion is balanced by the external 
truncation of the tidal field. The exact position at which this occurs will be dependent 
on the strength of the tidal field, and thus the galaxy model, and on the initial mass 
of the star cluster.

In the simulations presented here a star cluster orbiting the Galactic center at
50 kpc or more is free from disk shocking, given that the disk has an extent of 
$\sim$40 kpc. Disk shocking has demonstrated wounding effects on cluster stability 
(Gnedin \& Ostriker 1997; Vesperini \& Heggie 1997). In an effort to quantify the effect 
of disk shocking a full simulation was carried out with a star cluster in an orbit at 
10 kpc from the galactic center where the mass of the disk was placed in a central spheroid 
component instead of a disk. This showed that the presence of a disk will enhance the 
mass loss of the simulated cluster by 2.4$\times 10^3$ M$_{\odot}$ (13\% of the total mass) 
over a Hubble time, and will make its half-mass radius smaller by 0.6 pc (10\% of its size).
Thus at 10 kpc from the galactic center the effect of the disk in a star cluster, while 
subtle, is clearly measurable.

From the output of \texttt{NBODY6} we compute the half-mass relaxation time as:

\begin{equation}
t_{rh}=\frac{0.14N}{\ln\Lambda} \sqrt \frac{r_{hm}^3}{GM}
\label{eqn:eqnzz}
\end{equation}

where $N$ is the number of stars, $r_{hm}$ the half-mass radius, 
and $\Lambda = 0.4N$ the argument of the Coulomb logarithm (Spitzer \&
Hart 1971; Binney \& Tremaine 1987). Note that Giersz \& Heggie (1994) 
found a value of $\Lambda=0.11N$. The evolution of the relaxation time for 
each star cluster is homologous to the evolution of the half-mass radius 
with time shown in Figure 2. For all clusters it reaches a maximum value 
at the point when expansion driven by mass loss is equivalent to evaporation 
in the cluster (Gieles et al.\ 2011). The simulations carried out here show 
that this coincides with when the half-mass radius reaches its maximum as well. 
The number of relaxation times reached by each simulation after a 
Hubble time is given in Table 1, for those simulations that survive 
this long. Accentuated mass loss of the simulated 
star clusters accelerates their dynamical evolution. For example, 
the simulated cluster evolving at $R_{GC}=6$ kpc undergoes 90 relaxation 
times in 13 Gyr compared to only 3 for the simulated cluster at $R_{GC}=100$ kpc.


\subsection{Half-mass radius versus galactocentric distance at present time}

Figure 3 shows the half-mass radius of simulated star clusters orbiting at 
different galactocentric distances after 13 Gyr of evolution. The values used to 
create Figure 3 can be found in Table 1. After a Hubble time of evolution the 
half-mass radius defines a relationship with galactocentric distance which takes
the mathematical form of a hyperbolic tangent. This relation, plotted in 
Figure 3 as a solid line, is:

\begin{equation}
r_{hm}=r_{hm}^{max} \tanh (\alpha R_{GC})
\label{eqn:eqnn}
\end{equation}

where $r_{hm}$ is the half-mass radius and $r_{hm}^{max}$ is the maximum 
half-mass radius attained by simulated clusters evolving at 50 and 100 kpc
from the galactic center after a Hubble time- all radii in 3D. The maximum half-mass 
radius defines a plateau in the relationship between half-mass radius and 
galactocentric distance. Current relations defining the scale sizes of clusters 
as a function of galactocentric distance such as the empirical power-law 
$r_h=\sqrt{R_{GC}}$ (van den Bergh et al.\ 1991) do not include this flattening 
at large galactocentric distances. Note that the database used by van den Bergh 
et al.\ (1991) only included star clusters out to 32.8 kpc from the galactic center. 

The parameter $\alpha$ is a positive coefficient that defines the inner slope, i.e.\ 
within the inner 20 kpc. In this case its numerical value is 0.06. The parameter 
$\alpha$ is a proxy of the tidal field of the host galaxy that is in turn 
due to galaxy mass. Note that the onset of the plateau of Figure 3 at $R_{GC}\simeq 40$ 
kpc coincides with the approximate extent of the disk in our models. Beyond this 
distance globular clusters while remaining fully bound to the galaxy evolve 
in virtual isolation and are exempt from the truncating effects of the host 
galaxy tidal field. Therefore star clusters at 40 kpc and beyond have their sizes 
determined primarily by their internal dynamics (Spitzer \& Thuan 1972) 
and also to some extent by their initial size. 
Hurley \& Mackey (2010) showed that for clusters in a weak tidal field, differences 
in initial size can lead to long-term differences in half-mass radius, although 
the effect diminishes with age. 
Initial sizes of clusters are uncertain with some suggestions that the typical 
size-scale of a protocluster is $\sim 1\,$pc (Harris et al. 2010) but the 
actual size after gas removal (when the $N$-body simulations start) will depend 
on the star formation efficiency within the protocluster (Baumgardt \& Kroupa 2007). 
Thus it should be noted that the initial size of the cluster will have some 
bearing on the location of the plateau, i.e. $r_{\rm hm}^{\rm max}$. 


A set of simulations with a different initial mass was executed in order to 
establish a scaling relation that is independent of mass and will thus allow us
to make inferences towards more massive systems. Three full simulations, all of
them with identical initial conditions but with different initial number of 
particles and thus different initial masses were carried out at a galactocentric 
distance of $R_{GC}=8$ kpc. These three simulations have initial masses 
of $M=6.3 \times 10^4 M_{\odot}$,  $M=4.9 \times 10^4 M_{\odot}$, and
$M=3.2 \times 10^4 M_{\odot}$ with initial particle numbers of N=100 000, 75 000,
and 50 000 respectively. A unit-free relation that is interchangeable for these
three simulations with different initial masses is $M/M_{0}$ vs $t/t_{rh}$,
where $M_{0}$ is the initial cluster mass and $t_{rh}$ is the half-mass relaxation 
time (see also Baumgardt 2001). 

The relation above allows us to derive a predicted value of the half-mass radii 
after a Hubble time of evolution for simulations with twice the number of 
initial particles (i.e.\ $N$= 200 000). 
More massive clusters have longer relaxation times  $t_{rh}$ (Spitzer 1987). 
In fact, at any particular time, a star cluster with N= 200 000 has a  $t/t_{rh}$ smaller by a 
factor of 3/4 than the $t/t_{rh}$ of a star cluster with half the number 
of particles. We predict that after a Hubble time of evolution, the half-mass 
radii of clusters with $N$= 200 000 are equivalent to the half-mass radii 
of the simulations carried out here but at an earlier stage of their evolution.
The predicted half-mass radius values for models starting with $N$=200 000 are plotted 
in Figure 3 as blue stars. As this shows, 
the relation of half-mass radius vs.\ galactocentric distance postulated 
above (Eq. 5) holds true for models with higher masses. 


Hwang et al.\ (2011) derived the two-dimensional half-light radii of star 
clusters and extended star clusters in the halo of the dwarf galaxy NGC 6822. 
The main body of this dwarf galaxy has a scale size of 2.3 kpc (Billett et al.\ 2002).
The shape of the distribution of the effective radii of globular clusters from 
the center of NGC 6822 is strikingly similar to the half-mass radius vs. galactocentric 
distance relation shown in Figure 3. The distance at which the 
relation between half-light radius and galactocentric distance starts to plateau is 
about 4.5 kpc for NGC 6822.

In dwarf galaxies this $r_{hm}^{max}$ can be observed due to the smaller scales 
involved that can fit within the field of view of {\it HST} detectors. This 
raises the possibility of using the correlation between star cluster size and 
galactocentric radius to determine the structural parameters and physical 
characteristics of the different components of the host galaxy such as bulge 
mass, or disk scale-length.

The coincidence between Fig.\ 3 and the results of Hwang et al.\ (2011)
is encouraging, however, no information is available on the orbits of these 
clusters. A set of models taking into account the characteristics of NGC 6822 
will be needed before definite conclusions regarding the spatial distribution of 
the size of star clusters can be made.


\subsection{Core Radius}

The galactocentric distance of a star cluster also has an impact on its 
core radius, the onset of core collapse, and thus the ratio of core to 
half-mass radius. Figure 4 shows the evolution of the ratio of core to 
half-mass radius over 15 Gyr for simulations at a galactocentric 
distance of 4, 6, 10 and 100 kpc. Only selected simulations are plotted 
to preserve the clarity of the figure. The ratio $r_c/r_{hm}$ has been used 
in various ways to characterize the dynamical state of globular clusters, 
including as a possible indicator of the presence of an intermediate-mass 
black hole  (see Hurley (2007) for a discussion). 

The core radius, not directly depicted here, expands rapidly within 
the first 2 Gyr from $\sim2.6$ pc to $\sim4.1$ pc. While the expansion
of the core is similar at different galactocentric distances, the core to 
half-mass radius ratio is different for simulations evolving within
and beyond the inner 10 kpc of the galaxy. For simulations at 4 and 6 kpc 
the ratio $r_c/r_{hm}$ increases significantly during the initial 
5 Gyr of evolution. Accentuated mass loss at small galactocentric distances
also precipitates the onset of core collapse as seen in Figure 4. Core collapse
is reached at $\sim$ 7 Gyr for a star cluster at 4 kpc from the galactic center, 
just prior to dissolution, while core collapse is reached at 11 Gyr for a cluster 
at 6 kpc. Figure 4 shows that  clusters in orbits of 10 and 100 kpc do not 
reach the end of the core collapse phase in a Hubble time. This will continue
to be true for more massive clusters as the relaxation time will be even larger.



\subsection{The impact of orbital ellipticity}

As a test of the impact of orbital ellipticity on the size of star clusters
a simulation with a non-circular orbit was executed. This simulated star cluster 
has a perigalacticon of 4 kpc and an apogalacticon of 8 kpc, 
thus interposed between the two circular orbits at 4 and 8 kpc. The mass-loss rate
of the simulated star cluster evolving on an elliptical orbit closely resembles 
the mass-loss rate of the cluster evolving on a circular orbit at 8 kpc.
In fact, after a Hubble time of evolution the difference in mass between 
these two simulated clusters is only $\sim 10\%$. That is, the simulation
on a circular orbit at 8 kpc has $\sim 10\%$ more mass than the simulation on 
an elliptical orbit. Similarly, the half-mass radius is $\sim 12\%$ larger
for the circular simulation at 8 kpc after a Hubble time. 

King (1962) postulated that the tidal radius is imposed at perigalacticon
however this has been recently debated (K\"upper et al.\ 2010b, Baumgardt et al.\ 2010).
The exploratory findings presented above show that the half-mass radius
is closer to being set at apogalacticon. 


\subsection{Extended outer halo star clusters}

Mackey \& van den Bergh (2005) find a deficit of compact galactic young 
halo clusters at $R_{GC} > 40 $ kpc. With all clusters in this region 
having a larger than average effective radius, i.e. $r_h > 10$ pc, in agreement 
with the proposed relation of effective radius with galactocentric distance for star 
clusters $r_h \propto \sqrt{R_{GC}}$ (van den Bergh et al. 1991, McLaughlin 2000).

These extended clusters are believed to be accreted from now-destroyed satellite
dwarf galaxies with milder tidal fields (Mackey \& van den Bergh 2005). Under 
the initial conditions of the simulations carried out in this study, Figure 2 shows 
that there is no need for merger events with dwarf galaxies to grow extended star
clusters in the Milky Way at large galactocentric distances. 

The simulated star clusters evolving at 50 and 100 kpc from the galactic center 
undergo the initial expansion due to mass loss triggered by stellar 
evolution and then experience a small yet steady increase in half-mass radii (Figure 2). 
The tidal field is too weak at large distances from the galactic center to exert its 
truncating effects. Smaller initial sizes would lead to slightly smaller observed
sizes. However, increased initial masses would more than compensate for this effect
and still produce extended clusters at large galactocentric distances.


 \begin{figure} 
 \plotone{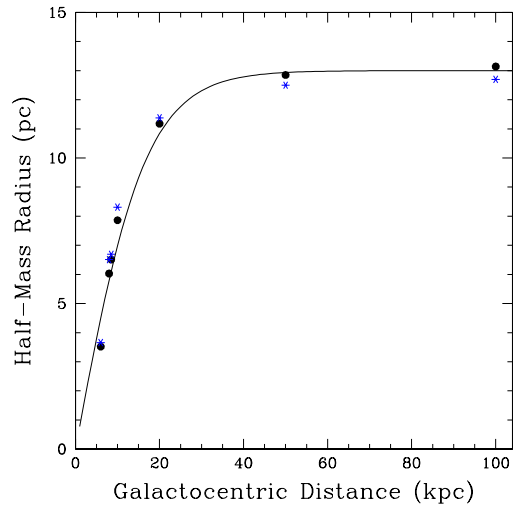} 
 \caption{3D half-mass radius vs. galactocentric distance of simulated star 
 clusters. Black dots give the half-mass radius of models at 13 Gyr. 
 The solid line is simply a hyperbolic tangent describing a new relation
 between half-mass vs.\ galactocentric distance. Blue stars depict the predicted value
 of the half-mass radius after a Hubble time for simulations with twice the
 initial number of particles and twice the initial mass. 
 \label{fig3}}
 \end{figure}



 \begin{figure} 
 \plotone{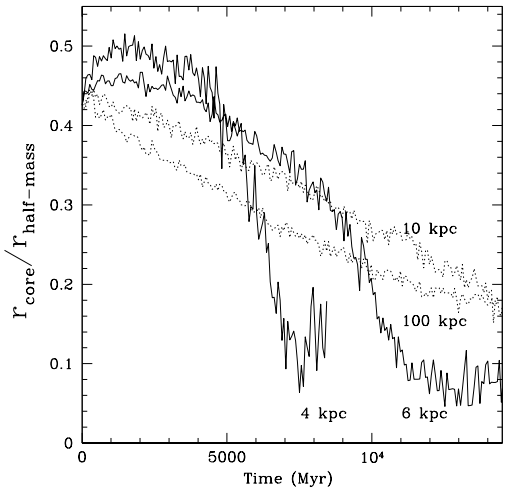} 
 \caption{Time evolution of the ratio of core over half-mass radius. Solid
 lines are models at 4 and 6 kpc while dotted lines are models at 10 and 100
 kpc from the galactic center. Both clusters with orbits at 4 and 6 kpc from 
 the galactic center reach core collapse before a Hubble time.
 \label{fig5}}
 \end{figure}



\subsection{A bimodal size distribution of star clusters orbiting dwarf galaxies}

Da Costa et al.\ (2009) find a bimodal distribution in the effective radius of 
globular clusters that are satellites of dwarf galaxies. Two peaks are seen
in the size distribution of star clusters: one at $\sim 3$ pc and a second peak
at $\sim 10$ pc. Da Costa et al.\ (2009) postulate that the second peak at 
10 pc is present when globular cluster systems evolve in a weak tidal 
field. They also note that extended star clusters in the Andromeda galaxy,
the Large Magellanic Cloud, and the Milky Way are found at large galactocentric 
distances. 

From the $N$-body models carried out for this work it can be seen that the 
first peak of the size distribution found by Da Costa et al. (2009) at 
$r_h\sim 3$ pc can be explained by clusters that are originally massive 
enough to survive the mass loss due to tidal stripping. Their size however 
is determined by tidal interactions with the host galaxy. The second peak 
in the distribution at $r_h\sim$ 10 pc would be created by star clusters evolving 
in a benign tidal environment, determined only by stellar evolution 
and internal dynamics.


\section{The dissolution of a star cluster}

Figures 1 and 2 show that the model cluster evolving on a circular orbit 
with a radius of 4 kpc around the galactic center does not survive a 
Hubble time. The fierce mass loss of a cluster evolving at 4 kpc from the galactic 
center drives the cluster to dissolution after 8.4 Gyr. This particular simulation 
was run twice with a different seed for the initial particle distribution to 
ensure that the dissolution before a Hubble time was reproducible. 

A simulation with a more dense and compact star cluster evolving at 4 kpc
from the galactic center was also computed. This simulation has an initial 
half-mass radius of 3.1 pc and the same mass of the other simulations presented
here, i.e.\ $M_{tot}\approx6.3 \times 10^4 M_{\odot}$. A higher density provides
shielding from the tidal field evident from this denser simulation surviving up  
13.3 Gyr. However it has less than one thousand stars remaining  
after 12.5 Gyr and is only left with $M_{tot}=764 M_{\odot}$.
This work provides a solid lower limit to the initial mass of star clusters 
seen today evolving at close range from the center of a Milky Way type galaxy.


The dissolution of the simulated star cluster at 4 kpc is real and 
of significance to explain the galactic globular cluster system radial 
density profile. The radial density distribution of Galactic globular clusters 
is described by a power law with a core (Djorgovski \& Meylan 1994). A simple power law
is a good fit for the radial distribution of globular clusters at large 
galactocentric distances. However, at $R_{GC}\le 3.5$ kpc, a flattening of the
radial distribution of clusters occurs making the overall distribution  
better described by a S\'ersic profile (S\'ersic, 1968) with a S\'ersic index 
of n=3 (Bica et al. 2006).


The central flattening of the globular cluster density profile has been 
explained as a result of a primordial density distribution with a depleted 
core (Parmentier \& Grebel 2005), incompleteness due to obscuration, or the 
result of enhanced destruction rates of globular clusters in the central 
regions of the galaxy. Near infrared surveys of the galactic core have revealed
few globular clusters during the last two decades (Dutra \& Bica 2000).
Incompleteness is thus a minor factor  when accounting for the apparent depletion 
of star clusters in the inner 3.5 kpc of the galaxy. Given that the simulation 
of a star cluster orbiting  at 4 kpc from the Galaxy core shows its complete 
dissolution before it reaches a Hubble time, this is an argument in favor 
of tidal disruption being responsible for an enhanced destruction rate of star 
clusters in the inner 4 kpc of the Galaxy, particularly those at the lower end
of the globular cluster mass function.


\begin{deluxetable}{ccccccc}
\tabletypesize{\footnotesize}
\tablecaption{Parameters of Simulated Star Clusters after a Hubble Time\label{tbl-1}} 

\tablewidth{0pt} \tablehead{
\colhead{$R_{GC}$} & \colhead{$r_{hm}$} & \colhead{Mass} & \colhead{$N$ stars}
&  \colhead{$N$ binaries} & \colhead{$t/t_{rh}$}
\\
\colhead{(kpc)} & \colhead{(pc)} & \colhead{$M_{\odot}$} & \colhead{  }
&  \colhead{  } & \colhead{  }
}
 
\startdata

4    &   --  & --         &   --   & --   & --   \\
6    & 3.5  & 3.1 $\times10^3$  & 4653   & 453  & 90.4 \\
8    & 6.0  & 1.3 $\times10^4$  & 25789  & 1703 & 17.9 \\
8.5  & 6.7  & 1.5 $\times10^4$  & 31602  & 1916 & 13.8 \\
10   & 7.9  & 1.9 $\times10^4$  & 44393  & 2331 & 9.1 \\
20   & 11.2 & 3.0 $\times10^4$  & 79591  & 3575 & 4.0 \\
50   & 12.9 & 3.5 $\times10^4$  & 94666  & 4069 & 3.0 \\
100  & 13.1 & 3.6 $\times10^4$  & 97506  & 4161 & 2.8 \\

 \enddata
 
\tablecomments{Parameters of simulated star clusters after a Hubble time.
 The simulated star cluster evolving on a 4 kpc orbit dissolves before a Hubble
 time. Column 1 gives the galactocentric distance of the circular orbit in which 
 the clusters evolve; Column 2: 3D half-mass radius; Column 3:  mass in solar masses; 
 Column 4:  number of stars; Column 5:  number of binaries; Column 6: number of half-mass
 relaxation times that have elapsed in a Hubble time (13 Gyr/$t_{rh}$). }

\end{deluxetable}



\section{Final remarks and future work}

The parametrization of the size scale of star clusters presented above can be used as a 
primary test of the radial distribution of extragalactic globular clusters with respect 
to the host galaxy in observational studies where spectroscopic information is not available. 
Globular clusters with  $r_h \gtrsim 10$ pc are expected to be at 
large galacocentric distances, extended clusters at $R_{GC} \lesssim 40$ kpc
can be expected to be artifacts of projection.  A well characterized size 
distribution of globular clusters across bulge, disk, and halo can also 
be used as an independent test of the mass of the different structural 
components of galaxies. 

The simulations with a realistic galaxy potential presented here yield, after a Hubble 
time of evolution, star clusters with the characteristics observed today. Further exploring
the initial values used for the set-up of the simulations is a natural follow-up to this work.
The initial values for the effective radius and concentration parameters 
can have an impact on the observed size distribution function (Harris et al.\ 2010).
We will also aim to make a comparison with the prescriptions of star cluster evolution
put forward by Alexander \& Gieles (2012).

New computing capabilities enabled by GPUs will allow simulations with
a larger initial number of star-particles. This should demonstrate that 
massive clusters can survive at small galactocentric distances, i.e. $R_{GC}=4$ kpc. 
Such simulations will improve our understanding of the initial size and mass distribution 
of globular clusters and test the theories that claim that globular clusters are the 
remnant nuclei of disk galaxies (B\"oker 2008).\\



\acknowledgments

We thank the anonymous referee for a detailed assessment of the 
original manuscript. This  research has made use  of the NASA Astrophysics 
Data System Bibliographic services (ADS), the NASA Extragalactic Database (NED), 
and  Google. Many thanks to Narae Hwang (NOAJ), Jeremy Webb (McMaster 
University), and Guido Moyano (Swinburne University) for enlightening 
discussions. Some of the simulations were carried out with gSTAR, 
a GPU based supercomputer hosted at Swinburne University in cooperation 
with Astronomy Australia Limited.




\end{document}